\documentclass{article}
\usepackage{epsfig}
\usepackage{amsmath,amssymb}

\usepackage{color}
\newcommand{\cO}[1]{{\textcolor{black}{#1}}}

\title{Flexible and robust networks}

 \author{S.~Vakulenko$^1$ and O.~Radulescu$^2$ \\
 \small  $^1$ Saint Petersburg State University of Technology and Design, St.Petersburg, Russia, \\
 \small  $^2$ DIMNP UMR CNRS 5235, University of Montpellier 2, Montpellier, France.}

\begin{document}



\maketitle



{\bf Abstract }
We consider networks with two types of nodes. The $v$-nodes, called centers,
are hyperconnected and interact one to another via many $u$-nodes, called satellites.
This centralized architecture, widespread in gene networks, possesses two fundamental properties.
Namely, this organization creates feedback loops that are capable to generate practically any prescribed patterning dynamics, chaotic or periodic, or having a number of equilibrium states. Moreover, this organization is robust with respect to random perturbations of the system.


\section{Introduction}

Flexibility and robustness are important properties of biological systems.  Flexibility means the
capacity to adapt with respect to changes of  environment whereas  robustness is the capacity
to support homeostasis in spite of environmental changes. Intriguingly, it seems that biological
systems could be in the same time robust and flexible. Development of an organism is robust to variations of initial conditions and environment, species can diversify in order to better
satisfy constraints imposed by a varying environment.

We discuss here flexibility and robustness problems for genetical networks
 of a special topological structure as a model for flexible and robust systems. In these networks,
highly connected hubs play the role of organizing centers (centralized networks). The hubs receive and dispatch  interactions. Each center
interacts with many weakly connected nodes (satellites). Similar ideas,
that such a "bow-tie" connectivity can play a role in robustness, have been proposed by (Zhao et al. 2006). In the field of random boolean networks (Kauffman 1969), the phase transitions from chaotic to frozen (robust) phases were related to scale-freeness and
heterogeneity of the network by (Aldana  2003).  We show that centralized network are capable to produce a number of patterns, while being protected against
environment fluctuations.

Network models  usually involve interactions between transcription factors
(TFs) (Reinitz et al. 1991).
In the last years, a great attention has been focused on microRNAs (He and Hannon 2004, Bartel 2009,
Hendrikscon et al. 2009, Ihui et al. 2010). MicroRNAs (miRNAs) are short ribonucleic acid (RNA) molecules, on average only 22 nucleotides long and are found in all eukaryotic cells. miRNAs are post-transcriptional regulators that bind to complementary sequences on target messenger RNA transcripts (mRNAs) and repress translation or trigger mRNA cleavage and degradation.
Thus, miRNAs have an impact on gene expression and
it was shown recently that they contribute to canalization of development (Li et al. 2009).
(Shalgi et al. 2007) shows the existence of many genes submitted to
extensive miRNA regulation with many TF among these "target hubs".
Without excluding other applications, we consider regulation of
TF by miRNAs as a possible example of a centralized network.
For this particular situation we generalize the TF network models (Reinitz et al. 1991)
to take into account miRNA satellites and the centralized architecture.
The interaction between network nodes is defined by
sigmoidal functions that can be defined by two parameters:
the maximum rates of production $r_i$, and sharpness constants $K_i$.
Other important parameters, play a key role, namely, the degradation
constants $\lambda_i >0$ of centers and satellites.

We obtain  a fundamental relation between the main network parameters. This relation ensures maximal robustness of the network with respect to random internal and external fluctuations, given a certain amount of flexibility defined as the number of attractors that are accessible to the network dynamics.
Our mathematical results have a transparent biological interpretation: centralized motifs can be simultaneously  flexible and robust. One can expect that miRNA molecules, being smaller with respect to TF, are more mobile and react faster to perturbations. This
property  plays a key role in the flexible and robust functioning of centralized motifs.

The paper is organized as follows.
Centralized networks are introduced in Section 2. We also formulate here an important assertion  on the flexibility of general centralized networks.
We show that these networks are capable to generate practically all  dynamics, chaotic or periodic, with any number of equilibrium states.
To study robustness with respect to random fluctuations,
in Section 3 we consider a toy model of simple centralized TF - miRNA networks with a single center. We show here, by an elementary way, that centralized networks with mutually repressive
hub-satellite interaction can produce many different robust patterns.


\section{Centralized networks}

Centralized networks have been empirically identified in molecular biology, where
the centers can be, for example, transcription factors, while the satellite
regulators can be small regulatory molecules such as microRNAs (Li et al. 2010).
Notice that, in the last decades, the theory of so-called scale-free networks has
become very popular.
Scale-free networks (Barabasi and Albert 2002, Lesne 2006) occur in many areas, in economics, biology and
sociology.
In the scale-free networks the probability $P(k)$ that a node is connected with $k$ neighbors,
has the asymptotics $Ck^{-\gamma}$, with $\gamma \in (2,3)$. Such  networks
typically contain a few strongly connected nodes and a number of satellite nodes. Hence,
scale-free networks are, in a sense, centralized.

In order to model dynamics of centralized networks we adapt a gene circuit model
 proposed to describe early stages of Drosophila (fruit-fly) morphogenesis
(Mjolness et al. 1991, Reinitz and Sharp 1995). To take into account the two types of the nodes,
 we use distinct variables $v_j$, $u_i$ for the centers and the satellites.
The real matrix entry
$A_{i j}$ defines the intensity of the action of
a center node $j$
on a satellite node $i$. This action can be either a repression $A_{i j} < 0$
or an activation $A_{i j} > 0$. Similarly, the matrices ${\bf B}$ and ${\bf C}$
define the action  of the centers on the satellites and the satellites on the centers, respectively. Let us assume
that a satellite can not act directly on another satellite
(the principle of {\em divide et impera}). We also assume that
satellites respond more rapidly to perturbations and are more diffusive/mobile
than the centers. Both these assumptions are natural if we identify satellites as microRNAs.

Let $M, N$ be positive integers, and let ${\bf A}, {\bf B}$ and ${\bf C}$ be matrices of the sizes $N \times M, M \times M $
and $M \times N$ respectively. We denote by ${\bf A}_i, {\bf B}_j$ and ${\bf C}_j$ the rows of these matrices.
To simplify formulas, we use the notation
$$
   \sum_{j=1}^M A_{ij} v_j = {\bf A}_i v, \quad \sum_{l=1}^M B_{jl} v_l= {\bf B}_j v, \quad
   \sum_{k=1}^N C_{jk} u_k={\bf C}_j u.
$$

Then, the {\em network model} reads (we exclude diffusion effects):
\begin{equation}
\frac{d u_i}{d t} =
\tilde r_i \sigma\left( {\bf A}_i v  - \tilde h_i\right) - \tilde \lambda_i u_i,
\label{cn1}
\end{equation}
\begin{equation}
\frac{d v_j}{d t} =
r_j \sigma \left({ \bf B}_j v + {\bf C}_j u
 - h_j\right) - \lambda_j v_j.
\label{cn2}
\end{equation}

We assume that the rate coefficients  $r_j, \tilde r_i$ are non-negative:
$ r_i, \tilde r_i \ge 0$.
Here  $i=1,..., N$, $j=1,...,M$ and
 $\sigma$ is a monotone and smooth (at least twice differentiable) {\em sigmoidal} function such
that
\begin{equation}
\sigma(-\infty)=0, \quad \sigma(+\infty)=1.
\label {eq2.5}
\end{equation}
 Typical examples can be given
by the Fermi  and Hill functions:
\begin{equation}
     \sigma(x)=\frac{1}{1 + \exp(- x)}, \quad \sigma_H(x)
     = \frac{x^p}{K_a^p  + x^p},
\label {eq2.6}
\end{equation}
where $K_a$, $p > 0$ are parameters and in the second case $x > 0$. For $x < 0$ we set $\sigma_H(x)=0$. Analytical
and computer simulation results are similar for both variants $\sigma$ and $\sigma_H$.


The parameters $\lambda_i, \tilde \lambda_i$ are degradation coefficients, and $h_i, \tilde h_i$ are thresholds
for activation.


Let us prove that the gene network dynamics defines a dissipative dynamics.
In fact, there exists an absorbing set ${\cal B}$ defined by
$$
{\cal B}=\{w=(u, v): 0 \le v_j \le r_j\lambda_j^{-1}, \ 0 \le u_i \le \tilde r_i \tilde \lambda_i^{-1}, \ j=1,...,M, \
 i=1, ..., N \}.
$$
 One can show, by comparison principles for ordinary differential equations, that
\begin{equation}
\begin{split}
 0 \le u_i(x, t) \le \tilde \phi_i(x) \exp(-\tilde \lambda_i t) + \tilde r_i\tilde \lambda_i^{-1}
 (1-\exp(-\tilde \lambda_i t)), \\
 0  \le v_i(x, t) \le \phi_i(x) \exp(-\lambda_i t) + r_i\lambda_i^{-1}(1-\exp(-\lambda_i t)).
 \end{split}
\label{est0}
\end{equation}
Therefore, solutions of (\ref{cn1}), (\ref{cn2})  exist for all times $t$ and they enters
for the set ${\cal B}$ at a time moment $t_0$ and then stays in this set for all $t > t_0$. So, our system defines a
dissipative dynamics and all concentrations are positive if they are positive at the initial moment. In mathematical terms, the Cauchy problem (initial value problem) for
our  system is well posed.

\section{Complex dynamics of centralized networks}

Let us show that the centralized networks  have a formidable power in dynamics generation.
First, we will find an asymptotic simplification of the dynamics, then show that
any dynamics, periodic, chaotic, or with a number of stable steady states
can be approximated by centralized networks.

\subsection{Simplified dynamics when satellites are fast}

We suppose here that the $u$-variables are fast and the $v$-ones are slow.
Then  the fast $u$ variables are slaved, for large times,
by the slow $v$ modes:  one has
 $u= U(v) + \tilde u$, where $\tilde u$ is a small correction. This means
 that, for large times, the satellite dynamics is defined
almost completely by the center dynamics.

To realize this approach,
let us assume that the parameters of the system satisfy the following conditions:
\begin{equation}
 \ |A_{jl}|, |B_{il}|, |C_{ij}|, |\tilde h_i|, |h_j| < C_0,
\label{cn3}
\end{equation}
where $i=1,2,...,N, \ \ i,l=1,...,M,\ j=1,...,N$,
\begin{equation}
 0 < C_1 < \tilde \lambda_j,
\label{cn4}
\end{equation}
and
\begin{equation}
 r_i=\kappa R_i, \quad \tilde r_i=\kappa \tilde R_i,
\label{cn41}
\end{equation}
where
\begin{equation}
 |R_i|, |\tilde  R_i|< C_5, \quad  \lambda_i=\kappa \bar \lambda_i, \ |\bar \lambda|  < C_6,
\label{cn41b}
\end{equation}
where $\kappa$ is a small parameter, and where all positive constants $C_k$ are independent
of $\kappa$.

\vspace{0.2cm}

{\bf Assertion 2.1.}
{\em
Under assumptions (\ref{cn3}), (\ref{cn4}),  (\ref{cn41})
 for sufficiently small $\kappa < \kappa_0$ solutions $(u, v)$ of (\ref{cn1}), (\ref{cn2})
  satisfy
\begin{equation}
u=U( v(t)) + \tilde u(t),
\label{cn5}
\end{equation}
where
the $j$-th component $U_j$ of $U$ is defined by
\begin{equation}
 - \tilde \lambda_j U_j= \kappa G_j(v),
\label{cn6}
\end{equation}
 where
$$
G_j=\tilde R_j
\sigma\left({\bf A}_{j} v( t)
 - \tilde h_j \right)
$$
The function $\tilde u$ satisfies estimates
\begin{equation}
|\tilde u|  < c\kappa^2 + R \exp(-
\beta t), \quad \beta > 0.
\label{cn8}
\end{equation}
The $v$ dynamics for large times $t > C_1 |\log \kappa |$ takes the form
\begin{equation}
\frac{d v_i}{d t} =\kappa F_i(u, v) + w_i,
\label{reddyn}
\end{equation}
 where $w_i$ satisfy
$$
|w_i| < c\kappa^2
$$
and
$$
F_i(u, v) =
R_i \sigma\left(
{\bf B}_{i} v + {\bf C}_{i} U(v)  - h_i \right) -\bar \lambda_i v_i.
$$
}

This assertion, known in computational biology as the quasi-steady state assumption, can be proved
by well known methods from the theory of differential equations (Henry 1981).

\subsection{Realization of prescribed dynamics by networks }

Our next goal is to show that dynamics (\ref{reddyn}) can realize, in a sense, arbitrary
 dynamics of the centers.
To precise this,
let us describe the method of realization of the vector fields
for dissipative systems (proposed by Pol\'a\v cik 1991,
for applications see, for example, Dancer - Pol\'a\v cik 1999, Rybakowski 1994, Vakulenko 2000).
This method is based on the   well developed theory of invariant and inertial manifolds,
see Marion 1989, Mane 1977, Constantin, Foias, Nicolaenko and Temam, 1989,
Chow-Lu 1988, Babin-Vishik 1988).
One can show that there are
systems  enjoying
the following properties:
\vspace{0.2cm}

{\bf A} {\sl These systems generate global
semiflows $ S_{\cal P}^t$
in an ambient  phase
space $H$. These semiflows depend on some
parameters $\cal P$ (which could be elements of another
parameter space $\cal B$).
They have global attractors
and finite dimensional local attracting invariant
$C^1$ (continuously differentiable) - manifolds $\cal M$, at least for some $\cal P$}.

{\bf B} {\sl
Dynamics of
$S^t_{\cal P}$ reduced
on these
invariant manifolds is, in a sense, "almost completely controllable".
It can be described as follows. Assume the differential
equations
\begin{equation}
\label{2.1}
\frac{dp}{dt}=F(p), \quad F \in C^1(B^n)
\end{equation}
define a dynamical system in the unit ball
${ B}^n \subset {\bf R}^n$.

For any prescribed dynamics (\ref{2.1}) and any $\delta >0$,
we can choose suitable parameters ${\cal P}={\cal
 P}(n, F, \delta)$ such that

{\bf B1} The semiflow $ S_{\cal P}^t$ has a $C^1$-
smooth locally attracting invariant manifold ${\cal M}_{\cal P}$
diffeomorphic to the ball ${B}^n$;

{\bf B2}
The reduced dynamics $ S_{\cal P}^t\vert_{{\cal M}_{\cal P}}$
is defined by equations
\begin{equation}
\frac{dp}{dt}=\tilde F(p, {\cal P}), \quad \tilde F \in C^1( B^n)
\label{tQ}
\end{equation}
where the estimate
\begin{equation}
|F -\tilde F|_{C^1({ B}^n)} < \delta
\label{est1}
\end{equation}
holds.
In other words, one can say that, by $\cal P$,
the  dynamics can be specified to
within an arbitrarily small error.}

Thus, all  dynamics
 can occur as inertial forms of these systems.  Such systems can be named {\em maximally
 dynamically flexible, or, for brevity, MDF systems}.

Such dynamics can be {\em chaotic}.
There is a rather wide broad in different definitions of
 "chaos". In principle, one can use here
any concept of chaos.
If this chaos is stable
under small $C^1$ -perturbations
this kind of chaos occurs in the dynamics of MDF
systems. To fix ideas, we  use here,
following Ruelle and Takens 1971, Newhouse, Ruelle and Takens 1971 Smale 1980, Anosov 1995), such a definition. We say that a finite
dimensional dynamics is chaotic if this
generates a non-quasiperiodic
hyperbolic invariant set $\Gamma$. If, moreover,
this set $\Gamma$ is attracting we say that $\Gamma$
is a chaotic (strange) attractor.
(For definition of hyperbolic sets,
see Ruelle 1989, Anosov 1995).
In this paper, we use only the following
basic property of hyperbolic sets,
  so-called Persistence (Ruelle 1989, Anosov 1995).
This means that the hyperbolic sets are, in a sense, stable(robust):
if (\ref{2.1}) generates the hyperbolic set $\Gamma$ and $\delta$
is sufficiently small, then dynamics (\ref{2.1}) also generates another
hyperbolic set $\tilde \Gamma$. Dynamics
(\ref{2.1}) and (\ref{tQ}) restricted to $\Gamma$
and $\tilde \Gamma$ respectively, are topologically orbitally equivalent
(on definition of this equivalence, see
Ruelle 1989, Anosov 1995).
It is important to mention that a chaos in dissipative systems may be
stable, in the sense of structural stability, and
although not yet observed in gene networks,
structurally stable chaotic itineracy is thought to play a functional role
in neuroscience (Rabinovitch 1998).

   Therefore, any possible chaotic robust dynamics
 can be generated by the MDF systems, for example,
the Smale horseshoes, Anosov
flows, the Ruelle-Takens-Newhouse chaos, see Newhouse, Ruelle, and Takens, 1971, Smale 1980, Ruelle 1989.
Some examples of the MDF systems were given
in Dancer- Pol\'a\v cik 1999, Rybakowski 1994, Vakulenko 2000.

Assertion 2.1  allows us to apply this approach to centralized network dynamics.
To this end, assume that (\ref{cn41}) and (\ref{cn41b})  hold. Moreover, let us assume
\begin{equation}
  \lambda_i=\kappa^2 \bar \lambda_i, \quad h_i=\kappa \bar h_i
\label{cbh1}
\end{equation}
where all coefficients $\bar h_i$
are uniform in $\kappa$ as $\kappa \to 0$.
We also assume that all
direct interactions between centers are absent, ${\bf B}={\bf 0}$.
This constraint is not essential.

Since $U_j=O(\kappa)$ for small $\kappa$, we can use the Taylor expansion
for $\sigma$ in (\ref{reddyn}). Then these equations reduce to
\begin{equation}
\frac{dv_i( \tau)}{d \tau} =
  \rho_i ( {\bf C}_{i} V( v)
  - \bar h_i) - \bar \lambda_i v_i +
\tilde w_i(t),
\label{cn11}
\end{equation}
where $\rho_i=\bar r_i \sigma'(0)$, $i=1,2,...,M$
and $\tau$ is a slow rescaling time: $\tau=\kappa^2 t$.
 Due to conditions (\ref{cbh1}), the corrections $\tilde w_i$ satisfy
$$
|\tilde w_i| < c\kappa.
$$

Let us focus now our attention to non-perturbed equation (\ref{cn11}) with $\tilde w_i =0$.
Let us fix the number of centers $M$. The number of satellites $N$ will be considered as a parameter.

The next important assertion immediately follows from well known approximation
theorems of the multilayered network theory, see, for example, Barron 1993, Funahashi and Nakamura 1993.
\vspace{0.2cm}

{\bf {Assertion 2.2.}}
{ \em Given a number $\delta > 0$, an integer $M$ and a vector field $F=(F_1, ..., F_M)$ defined
on the ball $B^M=\{ |v| \le 1 \}$, $F_i \in C^1(B^M)$,
 there are a number $N$, an $N\times M$ matrix ${\bf A}$, an $M \times N$ matrix ${\bf C}$
and coefficients $h_i$, where $i=1,2,...,N$, such that
\begin{equation}
|F_j(\cdot) - {\bf C}_{j} W( \cdot)|_{C^1(B^M)} < \delta,
\label{cn14}
\end{equation}
where
\begin{equation}
 W_i(v) = \sigma\left({\bf A}_{i} v
  - h_i \right),
\label{cn40}
\end{equation}
where $v=(v_1, ..., v_M) \in  {\bf R}^M$.
}

This assertion gives us a tool to control network dynamics.  Assume $\bar h_i=0$. Then equations (\ref{cn11}) with $\tilde w_i=0$
reduce to
the Hopfield-like equations for variables $v_i \equiv v_i(\tau)$ that depend only on $\tau$:
 \begin{equation}
\frac{ dv_l}{d \tau} =
    {\bf K}_{l} W(v)
 - \bar \lambda_l v_l,
\label{hop}
 \end{equation}
where $l=1,..., M$, the matrix $\bf K$ is defined by
$K_{lj}=\rho_l C_{lj} R_j \tilde \lambda_j^{-1}$. The parameters $\cal P$ of (\ref{hop}) are $\bf K$, $M$, $h_j$ and $\bar \lambda_j$.

In this case one can formulate the following result.
\vspace{0.2cm}

{{\bf {Assertion 2.3.}}
{\em
Let us consider a $C^1$-smooth vector field $Q(p)$ defined on
a ball $B_R \subset {\bf R}^M$ and directed strictly inside this ball at the boundary $\partial B^M$:
\begin{equation}
F(p) \cdot p < 0, \quad p \in \partial B^M.
\label{cn15}
\end{equation}
Then, for each $\delta > 0$, there is a choice of parameters $\cal P$ such that (\ref{hop}) $\delta$ -realizes
  system (\ref{2.1}). This means that (\ref{hop}) is a MDF system.
}

This  follows from Assertions 2.1 and 2.2.



\section{A toy model of centralized  network}

In this section we consider a simple centralized
network that, nonetheless,  can produce a number
of  point attractors (stable steady states). Due to its simple structure,
we can investigate here the robustness of this system.

Let us consider a central node interacting with many satellites.
This motif can appear as a subnetwork in a larger scale-free network.
In order to study robustness, we add  noise to the model. \cO{We consider
two types of stochastic perturbations.
The first type of perturbations is a Langevin type additive noise that
can simulate intrinsic stochastic fluctuations of gene expression
dynamics. The choice of additive noise is for the sake of simplicity,
however more general multiplicative noise can be used with no change
of the results. The second type of noise is a shot-like
perturbation that can simulate the external contributions to noise,
caused by the environment.}
Furthermore, we replace the sigmoid in \eqref{cn2} by a linear function.
This is justified in TF - miRNAs networks, where the action
of satellites (miRNA's) on centers (TF's) is post-transcriptional and produces
a modulation of the production rate of the center protein. This modulation
can be modeled by a soft sigmoid or even by a linear function.
Moreover, to simplify our model, we assume that all satellites are, in a sense,
equivalent.

The network
dynamics can be described then by the following equations:

\begin{equation}
\frac{d u_i}{d t} = -\lambda u_i + f_i(v) + \xi_i(t),
\label{Net1}
\end{equation}
\begin{equation}
\frac{dv}{dt} = -\nu v + Q(u) +\xi_0(t),
\label{Net11}
\end{equation}
where $f_i$, $Q$ are defined by
$$
Q(u) = a_0  + a \sum_{i=1}^n  u_i, \quad
f_i(u)=r \sigma(b (v - h_i)),
$$
Here $\xi_i$ are noises,
the coefficient $\lambda >0$ is a satellite mobility (degradation rate), $r >0$ is
the satellite maximum production rate, $b$ defines a sharpness of center action on the
satellites, $\nu >0$ is a center mobility (degradation rate), $a$ is the strength of the
satellites feedback action on
the center.

We consider the following type of noises:  non-correlated white noise
\begin{equation}
\langle \xi_i(t), \xi_j(t') \rangle = \beta_i \delta_{ij} \delta(t-t')
\label{SNet1}
\end{equation}
where $\beta_i>0$ are intensities, and shot-like noise
\begin{equation}
\xi_i(t)=\beta_i \eta_i \delta(t-\tau_j)
\label{SNet11}
\end{equation}
where $\tau_j$ are random shot times following a Poisson process, $\beta_i$ are noise amplitude coefficients,
and $\eta_i$ are random variables distributed uniformly on $[0,1]$. In numerical simulations we
set $\delta(t-\tau_j)=1$ with a probability $p_0 << 1$ and $\delta(t-\tau_j)=0$ with the probability $1-p_0$,
where $\tau_j=j\delta t$, $\delta t$ is a time step. Such
 noises $\xi_i$ can summarize the effect of a strong environment fluctuations
on the satellite and center expression.


 We study the problem under the following

{\bf Assumption. }{\em Let the derivatives of $f_i$ and $Q$ satisfy
$$
f_i^{'}(v) Q'(u) > 0 \ for \ all \ i, u, v.
$$}

Then one can show, following (Hirsch, 1988)  that the dynamics is monotone,
and, therefore, all trajectories converge to equilibria. The numerical simulations confirm this fact. Notice that the above assumption is not needed when satellites are fast, because
in this case the asymptotic dynamics is one dimensional and in dimension one all
the attractors are stable steady states (point attractors).
Although this simple system can not generate chaos or periodic behavior,
the number of point attractors can be arbitrarily large, and thus
this system is nonetheless flexible.

\subsection{Multistationarity of the toy model}

Let us  fix the signs of the satellite actions on the center assuming
that $a < 0$. This restriction is fulfilled in gene networks, where
the centers are transcription factors (TF) and the satellites are microRNAs
(indeed, usually microRNA can only repress transcription factors).
Let us show that the toy model
admits coexistence of any number of point attractors.

Let us make a transformation reducing (\ref{Net1}) and (\ref{Net11}) to a system of two equations
introducing a new variable $Z$ by
$$
Z=\sum_{j=1}^N  u_j,  \quad G(v)= \sum_{j=1}^N r \sigma(b(v- h_j)).
$$
Then, by summarizing eqs. (\ref{Net1}), one obtains
\begin{equation}
\frac{ dZ}{dt} =-\lambda Z + G(v),
\label{Net1Z}
\end{equation}
\begin{equation}
\frac{dv}{dt} = -\nu v + a_0  + aZ.
\label{Net2Z}
\end{equation}
This system is relatively simple and it can be studied analytically and numerically.
Since all
trajectories are convergent we obtain that the attractor consists of equilibria defined by
\begin{equation}
\lambda Z + F(Z)=0, \quad F(Z)=G(\nu^{-1}(a_0 + a Z)).
\label{Net3Z}
\end{equation}
Let ${\cal P}=\{b, h_i, N, a_0, a \}$ be free parameters that can be  adjusted.
Like to the previous section, we can "control" the nonlinearity $F$  by ${\cal P}$
and use the fact that $F(Z)$ can approximate
arbitrary smooth functions.
 The following
assertion  shows that the system is multi-stationarity with an
arbitrary number of point attractors:

\vspace{0.2cm}
{\bf Assertion 3.1.} {\em Let $N$ be a positive integer.
Then there are coefficients $ b, \lambda >0, r >0$, where $i=1,..., N$, $\nu >0$ and
$h_i, a_0, a$ in such a way that  equation (\ref{Net3Z})
has at least $n+1$ stable roots that can be placed in any given positions in the $Z$-space}.

The main idea of the proof  can be illustrated by Fig. 1
and holds in both cases of the Fermi and the Hill
sigmoids. Let us make a variable change
$w=\lambda Z/r$.
The steady states are solutions of the equation $r w = F(w)$, where
the function $F (w)$ is close to a step function with $N$ steps;
each step is given by the function $ \sigma( \gamma (w - \bar h_i) )$ that is close to Heaviside step function for large $\gamma$.
Here $\gamma$ is a  parameter that defines the sigmoid sharpness:
\begin{equation}
\gamma= a b r(\nu \lambda)^{-1}.
\label{gamma}
\end{equation}

The steady states of the system are given by the intersections
between the graph of $F(w)$ and the straight line of slope $r$.
An elementary argument shows that the intersections lying on
horizontal segments of the graph of $F(w)$ are stable attractors,
whereas the intersections on ascending vertical segments
correspond to repellers.

The position of the $i$-th
step in $w$-space is $\bar h_i$ and its height is $r$.
Under an appropriate choice of $\bar h_i$ this entails our assertion (see Fig. 1).
In the neural network theory,  $\gamma$ is known as {\em gain parameter}. This quantity, defined as the product of rates on sharpness divided on the product of degradation coefficients, gives
the maximal possible density of the equilibrium states in $w$-space.

It is useful to note that one gets $n+1$ attractors on the horizontal segments of the step function
provided that $\bar h_i$ decrease with $i$.

Notice that the main condition to obtain flexibility (multistationarity)
is the sharpness of the sigmoidal function, meaning that the gain parameter $\gamma$ should be large.
The construction is robust: we can vary $w_i, b, \bar h_i$ but the number of equilibria is conserved.


\subsection{Robustness and stability of attractors}

The roots of Eq.(\ref{Net3Z}) are point attractors  and then they are dynamically stable, otherwise, they are repellers and unstable. In Fig. 1, attractors
correspond to intersections of the straight line $y=rw$ with the curve $y=F(w)$,
lying on horizontal segments of the graph of $F$. A simple argument suggests
that the positions of these attractors are robust with respect to variations
of the thresholds $\bar h_i$.
Indeed, a perturbation of $\bar h_i$  induces a horizontal shift of the step
$\sigma( \gamma (w - \bar h_i) )$, and the
positions of the attractors are only slightly affected.

\begin{figure}[h!]
\centerline{
\includegraphics[width=80mm]{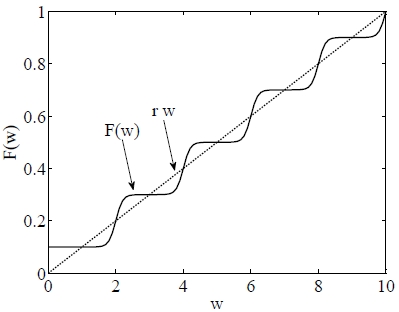}}
\caption{\small
 Intersections of the curve $y = F(w)$ and the straight line $y = rw$ correspond
 to steady states; intersections on horizontal segments of the graph of $F$ correspond to stable steady states.}
\label{fig1}
\end{figure}

More insight into robustness of the centralized toy model can be
obtained by considering the noisy case $\xi_i \ne 0$.

First, let us consider the case of  the Langevin noise (\ref{SNet1}).
We are interested in the robustness of the number and positions
of the attractors with respect to noises $\xi_i(t)$.
Near a point attractor, the equations (\ref{Net1Z}), (\ref{Net2Z})
can be linearized. The linearized dynamics is
defined by the following matrix ${\bf H}$:
\[ \left( \begin{array}{ccc}
-\lambda & \mu   \\
a & -\nu \\
\end{array} \right)\]
 where $\mu =G'(v_{eq})$.
For large $\gamma$ and for stable stationary states $\mu$ is small, $\mu=G'(v_{eq}) \to 0$ as $\gamma \to \infty$.
Let us assume that the noises $\xi_i$ are independent white noises.
Using standard results from the theory of linear
stochastic differential equations, see, for example, Keizer 1987,
it follows that
small deviations $\delta Z, \delta v$ from the equilibrium are normally
distributed with the density
\begin{equation}
  \rho(\delta Z, \delta v) = const \exp(-   X \cdot {\bf M}^{-1} \cdot X^{tr}), \quad X=(\delta Z, \delta v),
\label{gauss}
\end{equation}
where ${\bf M}$ is a symmetric, positively defined, $2 \times 2$
covariation  matrix with entries $m_{11}, m_{22},
m_{12}=m_{21}$. This matrix can be defined by the well known relation
(the fluctuation-dissipation theorem):
\begin{equation}
  {\bf H M +  M H}^{tr}  = - {\bf B},
\label{gauss2}
\end{equation}
 where ${\bf B}=diag(B_1^2, B_2^2)$ and, since the noises are non-correlated,
$B_1 =\sqrt{\sum_{i=1}^N  \beta_i^2}$, $B_2=\beta_0.$

As a result, a characteristic
fluctuation amplitude $F_A$ is proportional to the maximum  $\max\{\theta_1^{1/2}, \theta_2^{1/2}\}$
where $\theta_i$ are eigenvalues of ${\bf M}$.
Eq. (\ref{gauss2}) can be resolved explicitly  and $\theta_i$
can be found.

Now we can  investigate the following problem:
how to tune the parameters $\lambda, \nu$ and $a$ to obtain the minimal
fluctuation amplitude $F_A$ with respect to the
noise under a given  multistationarity level (this means $\gamma = \gamma_0 >> 1$ is  fixed but we can vary
the degradation rates $\nu, \lambda$).
 This optimization problem can be resolved numerically. The results,
which describe the optimal $\lambda_{opt}, \nu_{opt}$ as functions of $\gamma$,  are as follows.

The case {\bf A}), $B_2 >>  B_1$, the center is under a stronger noise than the satellites. Then the center degradation rate  $\nu_{opt}$ should be large, and $\lambda_{opt}$ is a small, decreasing in $\gamma$ function.

The case {\bf B}), $B_2 \le  B_1$, the center is under smaller noise than the satellites. Then the center
degradation rate  $\nu_{opt}$ should be smaller, $\lambda_{opt} > \nu_{opt}$, and the both parameters are decreasing in $\gamma$.
This situation is illustrated by Fig. 2.

\begin{figure}[h!]
\centerline{
\includegraphics[width=80mm]{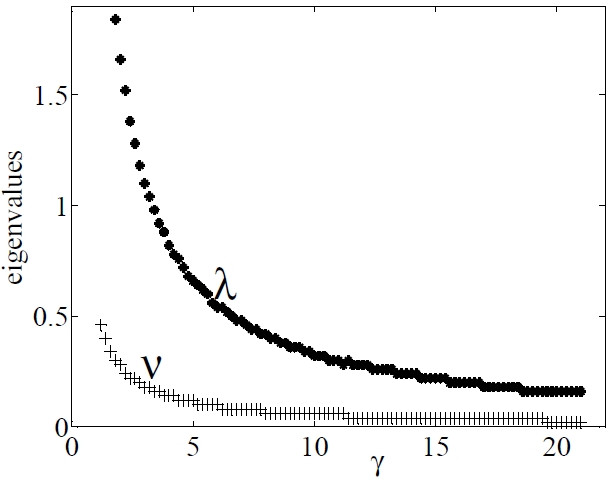}}
\caption{\small Optimal degradation parameters
$\nu_{opt}$, $\lambda_{opt}$ (minimizing the
eigenvalues of the matrix  ${\bf H}$)
as functions of the gain parameter $\gamma$ in the case $B_2 \le  B_1$, when the center is under smaller noise than the satellites.
}
\label{fig2}
\end{figure}

The classical  ideas of the invariant manifold theory, discussed in the preceding section, allow us
to systematize these results. The centralized network can function under two main and quite opposite
regimes. The first one arises when $\lambda >> \nu$. Then the satellite dynamics is slaved by the center motion. The center dominates and such a regime can be named {\em power of the center}.  This regime is stable if $\beta_0$ is small, but $\beta_i$ are large (the noises act on satellites mainly, case {\bf B}).  Considering that the noise intensity is larger for those components
that are expressed in larger copy numbers, the case should be representative for
miRNA-TF networks, when miRNA are in smaller copy numbers than the transcription factors.
In this case the noise perturb satellite states ($u_i$) but, since the satellites are controlled by the center state $v$,  satellites return to  the normal states
and dynamics is robust, the noise does not damage the attractor.
The opposite regime is when $\lambda <<  \nu$. Then, opposite to the previous situation,  the center dynamics is slaved by the satellites motion. Such a regime can be named {\em satellite democracy}.  This regime is stable when $\beta_0$ is large, but $\beta_i$ are small (the noise acts stronger on the center, case {\bf A}). Here the noise can perturb the center state ($v$) but this state can be restored by satellites. The large time dynamics is robust, again the noise does not damage the attractor.

Similar results are illustrated in the case of a shot noise in Fig. 3.

So, we obtain an interesting connection between robustness,
multistationarity
and network rate: to support robustness and multistationarity in a noisy
situation,
we should decrease the degradation constants. Multistationarity of
molecular
switches is important in decision making processes in differentiation,
development, and immune response of the organisms. Our finding means
that noise protected switches are necessarily slow.

\begin{figure}[h!]
\centerline{
\includegraphics[width=70mm]{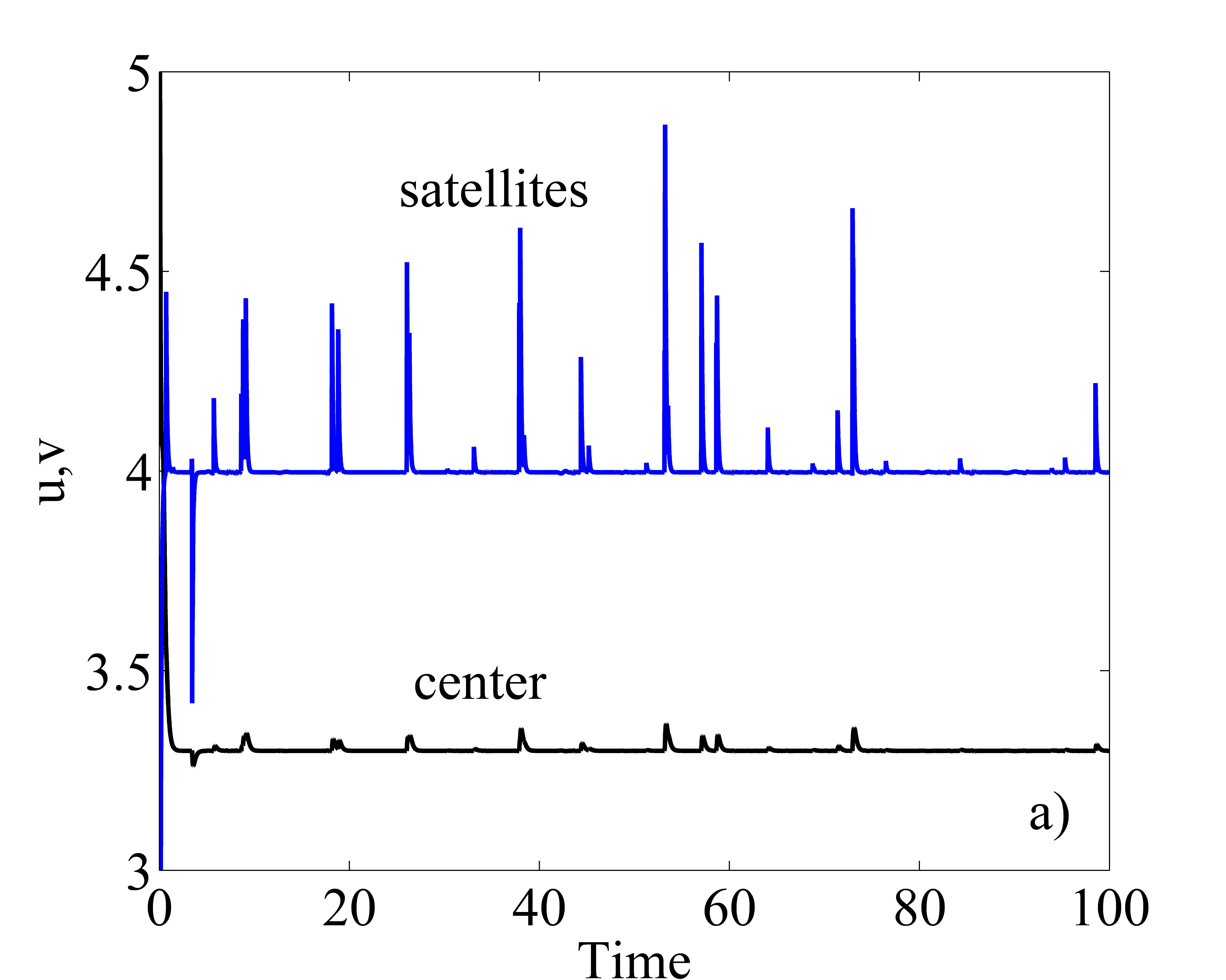}
\includegraphics[width=70mm]{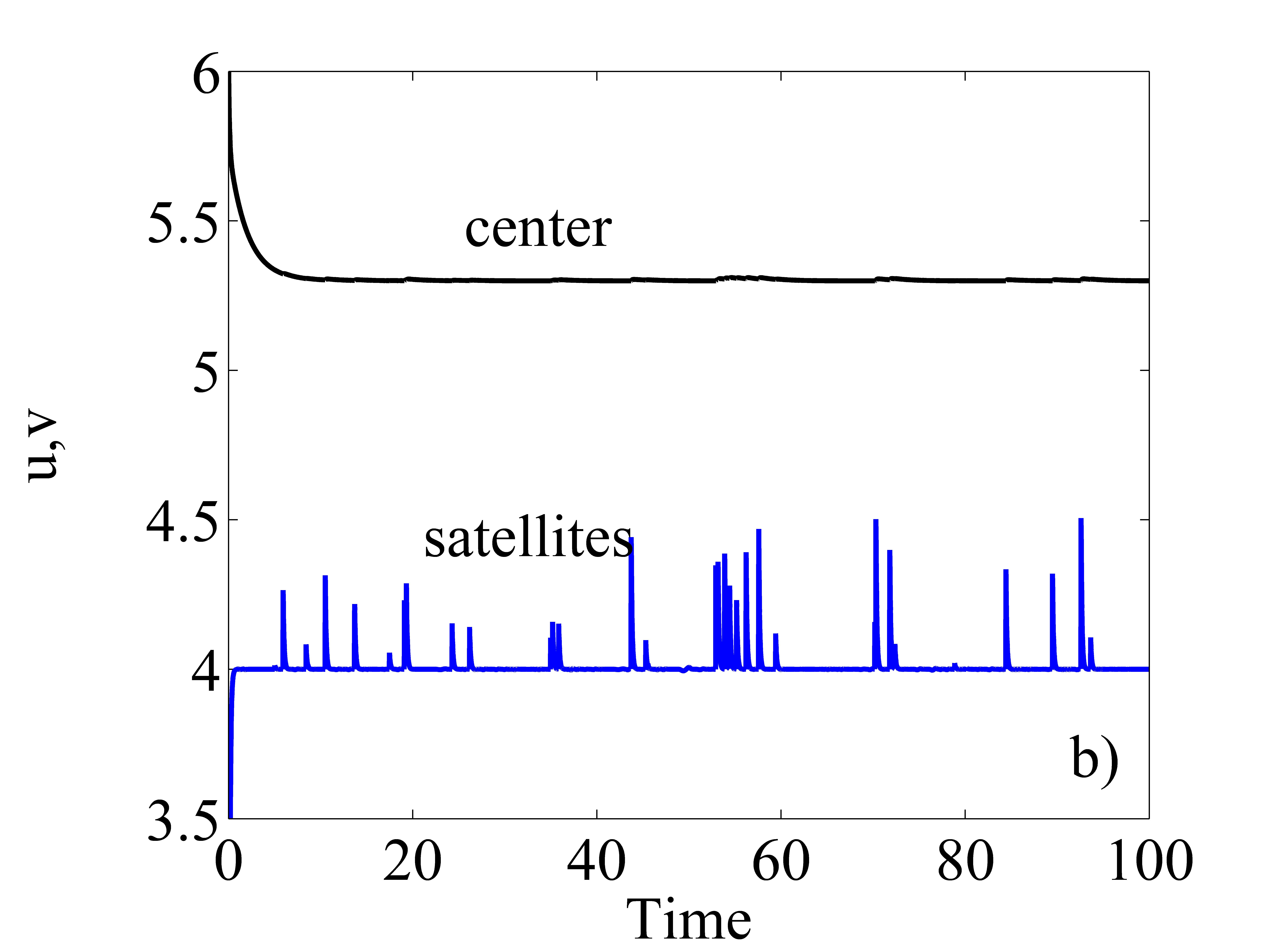}}
\caption{\small Numerical simulations of the system's trajectories under shot noise.
The parameters were as follows: $N=6$, $\sigma(z)=\sigma_H(z)$ with $p=4$, $b=20$, $K_a=1$, $h_i=i$, $r=1$, $\lambda=5$, $a_0=0.3\nu$, $x \in [0, 10]$ and $t \in [0, 120]$.
The parameter $\beta_i=0$ for all $i$ beside $i=3$, where $\beta_3=50$.
 The two functioning regimes correspond to different values
 of $\nu, a$, namely, $a=50,\   \nu=5$ (satellites democracy (SD), a)), and
 $a=5, \  \nu=0.5$ (power of the center (PC), b)).
 In the both cases the system shows multistationarity. For the chosen initial data
trajectories converge to an attractor $v \approx 5.3$ as $t >> 1$
in the PC regime and also in the SD regime
$v \to 3.3$. This means
that the fast center loses the attractor control,
while the slow center controls dynamics even under large deviations.
}
\label{fig3}
\end{figure}

\section{Conclusion}

We have considered networks with two types of nodes. The $v$-nodes, called centers,
are hyperconnected and interact one to another via many $u$-nodes, called satellites.
We show, by recently advanced  mathematical
methods, that this centralized network architecture, allows us to
control network dynamics to create complicated dynamical regimes.
This network organization creates feedback loops that are capable to generate practically all kinds of dynamics, chaotic or periodic, or having a number of equilibrium states. This strong flexibility could be crucial for
adaptive biological functions of these networks.

Using the simple example of a motif with a single center, we also argued
that centralized networks can perform trade-offs between flexibility and
robustness. To support both flexibility and robustness
in a noisy situation, the network should function
in a slow manner, i.e, we propose slow-down as a way to increase stability.

 Which of the nodes should be slowed-down depends on the fluctuations. Basic ideas
 from the invariant manifold theory show that if the noises act on the satellites,
 then, in order to conserve dynamics and the attractor structure, the center should
 be slow and controls the satellites (we called this regime
 {\em power of the center}). In the opposite case, when the noise acts on the center, the satellites should be slow in order to control the center and the global dynamics (we called this regime
 {\em satellites democracy}).

We did not consider here extrinsic noise or parametric variability of the
system, that we plan to study in the future.
We also think that the slow-down effect could be observed
in all systems where there is a separation into slow and fast variables, independently of architecture.




{\bf Acknowledgements}. The authors are grateful to Maria Samsonova and Vitaly Gursky for useful
discussions. We are thankful to M. S. Gelfand and his colleagues
for interesting discussions in Moscow.

The first author was supported by the Russian Foundation for Basic Research (Grant Nos. 10-01-
00627 s and 10-01-00814 a) and the CDRF NIH (Grant No. RR07801).
We are grateful to the anonymous referees for important remarks.



\end{document}